\documentclass[
aps,
pre,
numerical,
amsmath,
amssymb,
reprint
]{revtex4-2}

\usepackage{graphicx}
\usepackage{bm}
\usepackage{mathtools}
\usepackage{textcomp}
\usepackage{mathcomp}
\usepackage[english]{babel}
\usepackage{csquotes}

\begin{document}
\small

\title{Rescaled mode-coupling scheme for the quantitative description of experimentally observed colloid dynamics}

\author{Joel Diaz Maier}
\affiliation{Institut f\"ur Chemie, Universit\"at Rostock, 18051 Rostock, Germany}
\author{Joachim Wagner}
\email{joachim.wagner@uni-rostock.de}
\affiliation{Institut f\"ur Chemie, Universit\"at Rostock, 18051 Rostock, Germany}

\begin{abstract}
We describe experimentally observed collective dynamics in colloidal suspensions of model hard-sphere particles using a modified mode coupling theory (MCT). This rescaled MCT is capable to describe quantitatively the
wave-vector and time-dependent diffusion in these systems. 
Intermediate scattering functions of liquid-like structured dispersions are determined by means of static and dynamic light scattering experiments. The structure and short-time dynamics of the systems can be described quantitatively employing a multi-component Percus-Yevick ansatz for the partial structure factors and an effective, one-component description of hydrodynamic interactions based on the semi-analytical $\delta\gamma$-expansion. Combined with a recently proposed empirical modification of MCT in which memory functions are calculated using effective structure factors at rescaled number densities, the scheme is able to model the collective dynamics over the entire accessible time and wave-vector range and predicts the volume-fraction-dependence of long-time self-diffusion coefficients and the zero-shear viscosity
quantitatively. This highlights the potential of MCT as a practical tool for the quantitative analysis and prediction of experimental observations.
\end{abstract}

\maketitle

\section{Introduction}\label{scn:introduction}
Understanding the intricate connection between a material's macroscopic properties and the microscopic structure and dynamics of its constituents stands as a key endeavor in modern condensed matter physics. The quantitative prediction of a system's dynamics from its structural characteristics holds pivotal importance, serving as a fundamental step toward tailored material design \cite{vandergucht2018}.

Mode-coupling theory (MCT) is a widely recognized framework for calculating time-dependent correlation functions solely relying on static, time-independent inputs \cite{bengtzelius1984, szamel1991, janssen2018, szamel2013, reichman2005}. This theory, based on the Mori-Zwanzig projection-operator formalism \cite{zwanzig1960, zwanzig1961, mori1965, mori1965a}, explains the trapping of strongly interacting particles within nearest-neighbor cages as a vitrification mechanism which causes structural relaxation processes to occur on two distinct time-scales: The rapid $\beta$-relaxation is followed by the gradual $\alpha$-relaxation separated by an intermediate plateau. In strongly correlated fluids, arrested states are formed which are characterized by a diverging $\alpha$-relaxation time \cite{gotze1984}. MCT's significance lies not only in predicting an ideal glass transition but also in offering non-trivial descriptions of the transition between $\beta$- and $\alpha$-relaxation through characteristic power-laws and scaling relations \cite{gotze1985, gotze1990}. These predictions align well with empirical observations of glass-forming liquids \cite{kob1995, kob1995a, gotze1999}.

A recognized limitation of MCT is the quantitative disconnect of predicted glass transition temperatures or densities for various systems in comparison to experimental or computational results \cite{das2004}. MCT effectively reproduces the qualitative behavior of numerous dynamical observables such as coherent and incoherent density correlation functions \cite{weysser2010, voigtmann2003}, long-time collective- and self-diffusion coefficients \cite{banchio2000, banchio2018, nagele2002} or various rheological properties \cite{banchio1999, banchio1999a, priya2014, dicola2009, voigtmann2020, henrich2009}.
Achieving complete quantitative agreement, however, remains elusive. This discrepancy is attributed to MCT's tendency to overestimate correlations within the memory functions \cite{nauroth1997, flenner2005}. Notable counter-examples in this context are the correct quantitative prediction of the non-ergodicity parameter of liquid silica \cite{sciortino2001} and of hard spheres, either in bulk \cite{vanmegen1991, vanmegen1993} or in confinement \cite{jung2023}.

While it is evident that at least one of the underlying approximations of MCT (for a detailed derivation of the equations see e.g. Ref.~\onlinecite{reichman2005}) must be the reason for this exaggeration of correlations, precise insights have long been missing. Only very recently, \textcite{pihlajamaa2023} were able to dissect the effects of each separate approximation along the derivation of the MCT equations. Based on their findings, they identify routes to systematically improve standard MCT, but also highlight that many past attempts to extend MCT while upholding its foundational principles have not yielded wholly satisfactory results \cite{mayer2006, janssen2015, szamel2003, laudicina2022}.

With the systematic improvement of MCT still being an ongoing effort, in the meantime, developing simplified heuristic schemes for an improved MCT is valuable from a more practical perspective. Several past studies have proposed \emph{ad-hoc} solutions to match MCT calculations to experimental or computational outcomes, such as employing effective temperatures or packing fractions \cite{nauroth1997, flenner2005}, wave vector cutoffs or shifts \cite{nauroth1997}, or combinations thereof \cite{weysser2010, voigtmann2003, voigtmann2004}. Banchio \emph{et. al.} \cite{banchio1999, banchio1999a, banchio2000, banchio2018} found a rescaling of the density dependence of certain transport properties like viscosity or self-diffusion coefficient to be effective. Their approach not only accounts for MCT's limitations, but also for deviations induced by solvent-mediated hydrodynamic interactions (HIs), crucial in colloidal dispersion dynamics \cite{nagele1997}. Incorporating HIs directly into the MCT equations, while feasible in certain cases, remains a challenging endeavor \cite{nagele1998, nagele1998a, nagele1999}.

Recently, \textcite{amokrane2017} proposed a scheme in which the MCT equations are solved employing an effective structure factor as static input. By using a structure factor evaluated at lower density, the systematic overestimation of correlations in the memory function can approximately be compensated. Contrary to just using an overall effective temperature or density, this approach explicitly takes into account a possible wave-vector-dependence of the deviations introduced by the MCT approximations, beyond simply scaling the memory function by a factor. This method has been demonstrated to yield excellent agreement between MCT calculations and simulation data for various dynamic properties in uniform and binary hard-sphere fluids \cite{amokrane2019}.

In this work, we go beyond a comparison with computer simulations and apply this scheme directly to describe the experimentally observed collective dynamics of a model hard-sphere suspension. Direct comparisons between full, wave-vector-dependent MCT calculations and experimental data of density correlation functions have been reported \cite{vanmegen1993, voigtmann2003, gotze1999, banchio2018} but remain relatively rare compared to investigations involving simulations. Most investigations which interpret experiments in the scope of MCT either exploit universal scaling laws \cite{vanmegen1993, beck1999, mason1995} or rely on schematic, wave-vector-independent models \cite{krakoviack2002}, most often for quantities indirectly coupled to density relaxation as for example obtained from quasi-elastic scattering \cite{franosch1997, singh1998, ruffle1998, ruffle1999}, dielectric relaxation \cite{domschke2011} or rheological \cite{crassous2008, siebenburger2009, brader2010, diazmaier2022} experiments.

When working with real colloidal suspensions, both the particle size distribution and the HIs mediated by the surrounding medium need to be taken into account. While the explicit incorporation of continuous size distributions has gained popularity in both theoretical and simulation works \cite{weysser2010, zaccarelli2014, ninarello2017, pihlajamaa2023a, laudicina2022}, an accurate treatment of HIs, especially in polydisperse systems, remains challenging \cite{sierou2001, banchio2003, banchio2018, nagele1999}. 

It was recently demonstrated that the structure and short-time dynamics of dense, model hard-sphere suspensions consisting of silicone-stabilized poly(methyl methacrylate) (PMMA) particles can be modeled quantitatively  employing a multi-component Percus-Yevick ansatz in combination with an effective one-component treatment of HIs within the semi-analytical $\delta\gamma$-scheme \cite{diazmaier2024}. Based on this approach, we show the capability of MCT as a quantitative tool to describe the full, wave-vector- and time-dependent collective dynamics of this system and also test the predictive powers of this method by calculating self-diffusion coefficients and shear viscosities explicitly based on the accurate collective dynamics.

\section{Rescaled mode-coupling scheme}\label{scn:rescaled-mct}
Static structure factors $S(Q)$ and intermediate scattering functions $S(Q, t)$ (ISF) of model hard-sphere suspensions consisting of sterically stabilized PMMA particles are determined by means of static and dynamic light scattering experiments for a range of volume fractions in the liquid-like region. For details on the preparation and experimental procedures we refer to Ref.~\onlinecite{diazmaier2024}, where the here investigated suspensions were extensively characterized by means of static and dynamic light scattering. 

The optical properties of the studied particles are well-understood. Their scattering amplitude $b(Q)$ which for light scattering is essentially the Fourier transform of the refractive index distribution within a particle \cite{glatter2018} can be described by a core-shell model which also takes into account a slightly inhomogeneous distribution of the refractive index inside the core, induced by the partial penetration of the particles by the surrounding medium. This model of the scattering properties in combination with multi-component hard-sphere Percus-Yevick theory \cite{vrij1978, vrij1979, blum1979, blum1980} is able to describe quantitatively the scattered intensity of the suspensions over the whole accessible wave vector range.

The suspensions consist of particles with a mean radius $\langle R \rangle\approx285\,\mathrm{nm}$ which is a size well suited for light scattering experiments. It was shown that the effective hard-sphere radius of the particles somewhat decreases with increasing number density, most likely caused by changes in the chain conformation of the stabilizer chains when spheres come into close contact. This effect is accounted for during further analysis. The model fits additionally give access to the particle size distribution, which can be well described by a Schulz-Flory distribution \cite{schulz1939, flory1936} with probability density function
\begin{align} \label{eqn:schulz-flory}
	c(R) = \dfrac{1}{\Gamma(Z+1)} \left(\dfrac{Z+1}{\langle R \rangle}\right)^{Z+1} R^Z \exp\left(-\dfrac{Z+1}{\langle R \rangle} R\right),
\end{align}
where the dispersity $(\langle R^2 \rangle - \langle R \rangle^2)^{1/2}/\langle R \rangle = (Z+1)^{-1/2}$ amounts to approximately 7\,\% in our particular case. The different samples were assigned effective hard-sphere volume fractions in a range between $0.24\leq\varphi\leq0.52$ based on the fit results from the combined form factor and structure factor model. A volume fraction range still comparatively far from the glass transition at $\varphi\approx0.58$ \cite{vanmegen1991} constitutes a clean initial test case for the application of the proposed MCT scheme to analyze experimental results. Testing the same procedure in the deeply supercooled region close to the glass transition is the topic of ongoing investigation.

For our theoretical model, we consider a multi-component system comprising $n$ species of isotropically interacting, spherical particles. Following the method of \textcite{daguanno1992}, an $n$-component system representative for the continuous Schulz-Flory distribution is constructed, where already $n=5$ yields results which are indistinguishable from finer discretizations. The system is then specified by the radii $R_\alpha$ and number densities $\rho_\alpha$ of each species of identical particles. The total number density $\rho=\sum\rho_\alpha$, the number fractions $x_\alpha=\rho_\alpha/\rho$ and the total volume fraction $\varphi=4\pi/3 \sum \rho_\alpha R_\alpha^3$ can then be inferred from this information.

Photon correlation spectroscopy probes time-dependent fluctuations of the scattered intensity $I(Q, t)$ from which the normalized, measurable ISF $\phi_{\mathrm{M}}(Q, t) = S_{\mathrm{M}}(Q, t) / S_{\mathrm{M}}(Q)$ is determined \cite{Berne1976}. $\phi_{\mathrm{M}}(Q, t)$ is also termed density correlation function (DCF), since it is essentially the autocorrelation function of the Fourier components of the microscopic density \cite{janssen2018}. We employ the descriptor \enquote{measurable} to emphasize that observables directly derived from scattering experiments always in some form involve an average over the scattering amplitudes $b(Q)$ and do not constitute simple thermodynamic averages like when accessed for example by computer simulations. The static structure given by the partial structure factors $S_{\alpha\beta}(Q)$, the composition expressed by the number fraction $x_\alpha$ and the scattering amplitudes $b_\alpha(Q)$ of the suspension are considered to be known quantities for our multi-component system during further analysis. With this in mind, the time dependence of the measurable ISF
\begin{align}\label{eqn:measurable-isf}
	S_{\mathrm{M}}(Q, t) = \dfrac{\sum\limits_{\alpha \beta} (x_\alpha x_\beta)^{1/2} \, b_\alpha(Q) \, b_\beta(Q) \, S_{\alpha \beta}(Q, t)}{\sum\limits_{\alpha} x_\alpha \, b^2_\alpha(Q)},
\end{align}
is given by a weighted average of the elements $S_{\alpha \beta}(Q, t)$ of the matrix of partial ISFs $\mathbf{S}(Q, t)$ \cite{hansen1991}. Note that the partial structure factors $S_{\alpha \beta}(Q)$ in our convention follow the property $S_{\alpha \beta}(Q\to\infty) = \delta_{\alpha\beta}$, with $\delta_{\alpha\beta}$ being the Kronecker symbol. 

The problem of calculating $S_{\mathrm{M}}(Q, t)$ is thus shifted to computing $\mathbf{S}(Q, t)$. Within MCT, the time evolution of $\mathbf{S}(Q, t)$ is, neglecting HIs, determined by \cite{nagele1999} 
\begin{widetext}
\begin{align}\label{eqn:mct-collective}
    \dfrac{\partial}{\partial t} \mathbf{S}(Q, t) &+ Q^2 \mathbf{D} \mathbf{S}^{-1}(Q) \mathbf{S}(Q, t) + \mathbf{D}\int\limits_0^t \mathbf{M}(Q, t-t')\dfrac{\partial}{\partial t'} \mathbf{S}(Q, t') \mathrm{d}t' = 0,
\end{align}
where $\mathbf{M}$(Q, t) is the matrix of irreducible memory functions with elements
\begin{align}\label{eqn:mct-collective-memory}
    M_{\alpha\beta}(Q, t) = \dfrac{1}{16\pi^3 \left(\rho_\alpha \rho_\beta\right)^{1/2}}\sum_{\gamma\gamma'\delta\delta'} \int V_{\alpha\gamma\delta}(\mathbf{Q}, \mathbf{k}) V_{\beta\gamma'\delta'}(\mathbf{Q}, \mathbf{k}) S_{\gamma\gamma'}(k, t) S_{\delta\delta'}(|\mathbf{Q}-\mathbf{k}|, t) \mathrm{d}\mathbf{k}.
\end{align}
The vertices are given by
\begin{align}\label{eqn:mct-collective-vertex}
    V_{\alpha\gamma\delta}(\mathbf{Q}, \mathbf{k}) = \dfrac{\mathbf{Q}\cdot\mathbf{k}}{Q} \delta_{\alpha\delta} C_{\alpha\gamma}(k) + \dfrac{\mathbf{Q}\cdot(\mathbf{Q}-\mathbf{k})}{Q} \delta_{\alpha\gamma} C_{\alpha\delta}(|\mathbf{Q}-\mathbf{k}|),
\end{align}
\end{widetext}
with the dimensionless, weighted partial direct correlation functions $C_{\alpha\beta}(Q) = (\rho_\alpha\rho_\beta)^{1/2} c_{\alpha\beta}(Q) = \delta_{\alpha\beta} - (S^{-1})_{\alpha\beta}(Q)$ which are related to the partial structure factors by the multi-component Ornstein-Zernike relation \cite{hansen2013}. $\mathbf{D}$ is a diagonal matrix of Stokes-Einstein diffusion coefficients of species $\alpha$.

In a single-component system, hydrodynamic interactions lead to a rescaled short-time diffusion coefficient 
\begin{align}
D_{\rm eff}(Q) &= -\dfrac{1}{Q^2}\lim\limits_{t\to "0"} \dfrac{\partial}{\partial t} S(Q,t) = H(Q)\dfrac{D_0}{S(Q)}\,
\end{align}
where the notation $t\to "0"$ indicates a short time limit of structural relaxation times beyond momentum relaxation. The dimensionless scaling factor $H(Q)$ in the resulting extended de Gennes relation is the hydrodynamic function \footnote{We use here a slightly different notation with $H(Q)$ as a dimensionless scaling factor, whereas in \cite{nagele1999} $H'(Q)=H(Q) D$ with the dimension of a diffusion coefficient is used.}. Based on ideas by \textcite{medina-noyola1988} and \textcite{brady1993, brady1994}, recently recompiled by \textcite{riest2015}, it is an adequate approximation to assume that HIs in hard-sphere systems mostly influence the short-time behavior of any relaxation process, which is strengthened by the observation that for many transport properties, it is sufficient to rescale theoretical outcomes of non-HI methods with HI-included short-time/high-frequency contributions to quantitatively match experimental data \cite{banchio2000, banchio1999}. The hydrodynamic function of hard-sphere dispersions can in principle be calculated via computationally expensive accelerated Stokesian dynamics simulations \cite{sierou2001, banchio2003} or the approximate $\delta\gamma$-scheme introduced by \textcite{beenakker1983, beenakker1984}. The latter is restricted to uniform dispersions but offers the advantage of a semi-analytical treatment.

The collective short-time diffusion of the here presented experimental system was thoroughly investigated in Ref.~\onlinecite{diazmaier2024}. Therein, it was discussed that the measurable hydrodynamic function $H_\mathrm{M}(Q)$, acquired from quasi-elastic scattering experiments, can be well described by the effective one-component formulation
\begin{align} \label{eqn:hydrodynamic-function}
    H_\mathrm{M}(Q) = H_\mathrm s + H_\mathrm d(Q) = H_\mathrm s + A H_\mathrm{d}^{\delta\gamma}(Q^*),
\end{align}
with $Q^* = \alpha (Q-Q_\mathrm{m})$ where $Q_\mathrm{m}$ is the wave vector at the principal peak's location. $H_\mathrm{M}(Q)$ separates into a wave-vector-independent self part $H_\mathrm s$ and a wave-vector-dependent distinct part $H_\mathrm d(Q)$. The function $H_\mathrm{d}^{\delta\gamma}(Q)$ is the distinct part calculated within the one-component $\delta\gamma$-scheme \cite{beenakker1983, beenakker1984}, supplied with the size-average of the partial structure factors $\langle S(Q) \rangle$ as static input. The self-part $H_\mathrm s$ and the empirical parameters $A$ and $\alpha$ are treated as fit parameters to match the experimental data. 

\begin{table}[b]
	\caption{\label{tab:volume-fractions} Hard-sphere volume fractions $\varphi$ for all suspensions investigated and optimum rescaled volume fractions $\varphi^\ast$ used to describe experimental data. The relation between $\varphi$ and $\varphi^*$ can be conveniently expressed either via a scaling factor $\alpha=\varphi^*/\varphi$ or a shift $\Delta\varphi = \varphi - \varphi^*$.}
	\begin{ruledtabular}
		\begin{tabular}{cccc}
			$\varphi$ & $\varphi*$ & $\alpha=(\varphi-\Delta\varphi) / \varphi$ & $\Delta\varphi = (1-\alpha) \varphi$\\
			\hline
			0.52 & 0.447 & 0.86 & 0.073\\
			0.50 & 0.420 & 0.84 & 0.080\\
			0.48 & 0.393 & 0.82 & 0.087\\
			0.47 & 0.381 & 0.81 & 0.089\\
			0.44 & 0.349 & 0.79 & 0.091\\
			0.38 & 0.292 & 0.77 & 0.088\\
			0.33 & 0.250 & 0.76 & 0.080\\
			0.24 & 0.173 & 0.72 & 0.067
		\end{tabular}
	\end{ruledtabular}
\end{table}

To include this effective one-component formulation in the MCT scheme of a multi-component system, the time-dependence of the dynamic structure factor needs to be rescaled by a matrix $\mathbf{H}_{\rm eff}(Q)$ with effective hydrodynamic functions
\begin{align}
    [H_\mathrm{eff}(Q)]_{\alpha\beta} = \delta_{\alpha\beta} H_\mathrm s + (x_\alpha x_\beta)^{1/2} H_\mathrm d(Q)
\end{align}
as elements. Herewith, Eq. \eqref{eqn:mct-collective} reads as
\begin{align}\label{eqn:mct-collective-modified}
\dfrac{\partial}{\partial t} \mathbf{S}(Q, t) + Q^2 \mathbf{H}_\mathrm{eff}(Q)\mathbf{D} \mathbf{S}^{-1}(Q) \mathbf{S}(Q, t) &\notag \\
 + \mathbf{H}_\mathrm{eff}(Q)\mathbf{D}\int\limits_0^t \mathbf{M}(Q, t-t')  \dfrac{\partial}{\partial t'} \mathbf{S}(Q, t') \mathrm{d}t'& = 0\,.
\end{align}
This approximation neglects the influence of the scattering amplitudes $b(Q)$ on $H_\mathrm{M}(Q)$ which, however, only causes a very small disturbance of the overall measurable relaxation rate, not noticeable within experimental uncertainty: Since size-dispersity effects were found to mostly influence the potential part of the interparticle interactions in the systems studied here \cite{diazmaier2024}, the differences in scattering power of differently sized species are predominantly mediated by the structure factors and not by the hydrodynamic functions. With the scheme constructed in this way, the short-time limit of the resulting measurable ISF $S_\mathrm M(Q, t)$, calculated via Eq.~\eqref{eqn:measurable-isf}, matches the experimental data from Ref.~\onlinecite{diazmaier2024}. 

The empirical modifications proposed by \textcite{amokrane2017} can now be incorporated into the memory function [Eq.~\eqref{eqn:mct-collective-memory}]. For this, we first define a scaled volume fraction $\varphi^* = \alpha \varphi = \varphi - \Delta\varphi$ which is connected to the original volume fraction $\varphi$ either via a scaling factor $\alpha$ or equivalently via a shift $\Delta\varphi$ and later needs to be determined by matching the MCT outcome to the experimental data. Now, a set of effective partial structure factors $S_{\alpha\beta}^*(Q)$ is calculated with $\varphi^*$ instead of $\varphi$ as an input parameter and subsequently, every occurrence of the weighted partial direct correlation functions $C_{\alpha\beta}(Q)$ in the vertex [Eq.~\eqref{eqn:mct-collective-vertex}] is replaced by $C^*_{\alpha\beta}(Q) = \delta_{\alpha\beta} - (S^{-1})^*_{\alpha\beta}(Q)$.
\section{Collective Dynamics}
\begin{figure}[b]
	\includegraphics[width=1\linewidth]{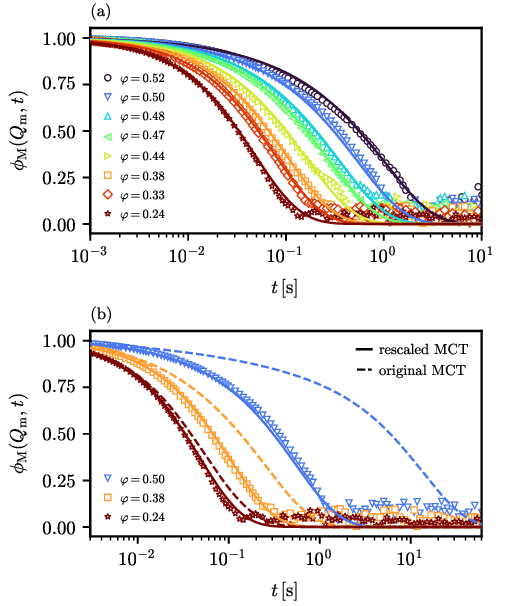}
	\caption{\label{fig:mct-exp} a) Time dependence of the measurable density correlation function $\phi_\mathrm M (Q_\mathrm m, t)$ at the principal structure factor peak's location $Q_\mathrm m$ for volume fractions as indicated in the legend. The solid lines are the results of the rescaled mode-coupling scheme described in the text. b) Comparison between $\phi_\mathrm M (Q_\mathrm m, t)$ calculated with rescaled volume fractions and results from the original MCT with an unmodified memory function.}
\end{figure}
The MCT equations for the presented scheme are numerically solved employing common methods \cite{fuchs1991, flenner2005, pihlajamaa2023b}. From the resulting partial ISFs $S_{\alpha\beta}(Q, t)$, the measurable ISF $S_\mathrm M(Q, t)$ is then calculated according to Eq.~\eqref{eqn:measurable-isf}. The optimum scaled volume fraction $\varphi^*$ used in the model was evaluated by minimizing deviations between experimental data and theory in the whole accessible time and wave-vector range. Optimum rescaled volume fractions $\varphi^*$ and actual volume fractions $\varphi$ are compared in Table~\ref{tab:volume-fractions}. 

\begin{figure}[b]
	\includegraphics[width=1\linewidth]{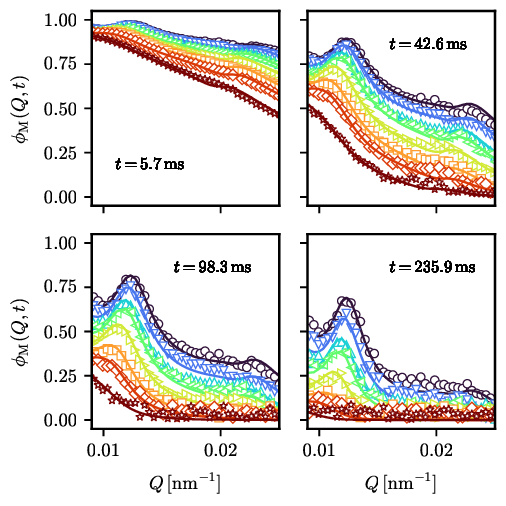}
	\caption{\label{fig:mct-exp-Q-dependence} Wave vector dependence of the measurable density correlation function $\phi_\mathrm M (Q, t)$ for chosen waiting times as indicated. The symbols are the same as in Fig.~\ref{fig:mct-exp}, with the solid lines being results of the rescaled mode-coupling scheme.}
\end{figure}

Fig.~\ref{fig:mct-exp}(a) shows a comparison of the time dependence of experimentally obtained measurable DCFs $\phi_\mathrm M (Q_\mathrm m, t)$, evaluated at the wave vector $Q_\mathrm m$ corresponding to the principal structure-factor peak, and theoretical results according to the rescaled mode-coupling scheme. In conjunction, in Fig.~\ref{fig:mct-exp-Q-dependence}, the wave-vector dependence of the same correlation functions is displayed for several delay times. For all investigated volume fractions, the experimental relaxation functions can be described quantitatively by the theory over almost the whole accessible time and wave-vector range. As expected from liquid-like systems relatively far from the glass transition, the relaxation is profoundly stretched for the larger volume fractions, but a separation into two different relaxation regimes is not yet observable. The rescaled theory captures this stretched decay accurately. While some minor deviations are noticeable, these can be well explained by statistical fluctuations of the experimental values, most apparent at long delay times where $\phi_\mathrm M (Q_\mathrm m, t) \lesssim 0.2$. The theoretical prediction itself is technically also subjected to uncertainties from the input parameters propagated through the scheme, however, the precise estimation of an uncertainty range is difficult due to the complexity of the model. 

Fig.~\ref{fig:mct-exp}(b) additionally displays for chosen densities a comparison between the predictions of the MCT scheme with and without scaling factors for the volume fraction. It is evident, especially at larger volume fractions, that standard MCT significantly overestimates the structural relaxation time. Clearly, this emphasizes the considerable improvements enabled by this comparatively simple rescaling approach, especially considering that the experimental values are matched not only for a conveniently chosen wave vector, but for all wave vectors simultaneously.  

\section{Prediction of Transport Properties}

Beyond the successful quantitative description of the experimentally observed collective relaxation, we test if related transport properties can be predicted correctly based on the parameterized scheme without further adjustable parameters. 

\begin{figure}[b]
	\includegraphics[width=1\linewidth]{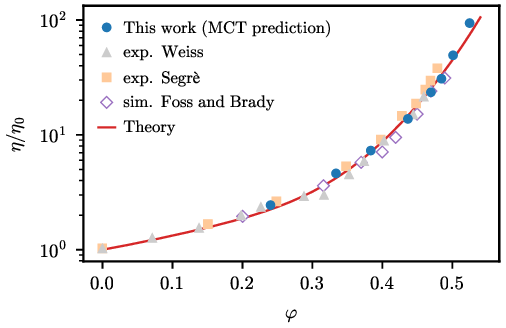}
	\caption{\label{fig:viscosity} Reduced zero-shear viscosity $\eta / \eta_0$, as predicted by the rescaled mode-coupling scheme, compared to experimental data by \protect\textcite{segre1995} and \protect\textcite{weiss1998} and results from Stokesian dynamics simulations by \protect\textcite{foss2000}. Additionally, a theoretical prediction based on a modified scaling expression proposed by Brady [Eq. 56 in Ref.~\protect\onlinecite{riest2015}] is given by the solid line.  
	}
\end{figure}

The shear viscosity $\eta$ in the limit of small shear rates can be directly calculated within MCT from the partial ISFs and is given by \cite{nagele1998a}
\begin{align}\label{eqn:viscosity}
    \dfrac{\eta - \eta_\infty}{\eta_0}  = \dfrac{k_\mathrm B T}{60\pi^2} \int\limits_0^\infty \int\limits_0^\infty Q^4 \, \mathrm{Tr}\left[ \left( \dfrac{\mathrm d \mathbf{C}(Q)}{\mathrm d Q} \cdot \mathbf{S}(Q, t) \right)^2 \right] \,\mathrm d Q \, \mathrm d t,
\end{align}
with $\eta_0$ being the viscosity of the surrounding medium and $\eta_\infty$ being the hydrodynamic high-frequency contribution. We use the accurate parameterization given by Eq.~36 in Ref.~\onlinecite{riest2015} to calculate $\eta_\infty$. The resulting predictions for $\eta/\eta_0$ in dependence on the volume fraction $\varphi$ are displayed in Fig.~\ref{fig:viscosity} where they are compared to independent results of experimental \cite{segre1995, weiss1998} and simulation studies \cite{foss2000} from the literature along with a theoretical prediction based on a generalized Stokes-Einstein relation \cite{riest2015}. Strikingly, the values predicted from the collective dynamics fit nicely among these results which clearly shows that if a mode-coupling scheme can correctly describe the collective density relaxation of a realistic suspension, the viscosity calculated via Eq.~\eqref{eqn:viscosity} is consistently also quantitatively correct. 

The self-dynamics of the system are also accessible within MCT. This however, requires the self-consistent solution of a supplementary set of equations. For the self-ISFs $S^s_\alpha(Q, t)$, these are given by \cite{nagele1997}
\begin{widetext}
\begin{align}\label{eqn:mct-self}
    \dfrac{\partial}{\partial t} S^s_\alpha(Q, t) + Q^2 H_s D^0_\alpha S^s_\alpha(Q, t) + H_s D^0_\alpha\int_0^t M^s_\alpha(Q, t-t') \dfrac{\partial}{\partial t'} S^s_\alpha(Q, t') \,\mathrm{d}t' = 0,
\end{align}
with the memory-kernels
\begin{align}
    M^s_\alpha(Q, t-t') = \dfrac{1}{8\pi^3 \rho_\alpha} \int \left(\dfrac{\mathbf{Q}\cdot\mathbf{k}}{Q}\right)^2 S^s_\alpha(|\mathbf{Q}-\mathbf{k}|, t) \sum_{\delta\delta'} C_{\alpha\delta}(k) C_{\alpha\delta'}(k) S_{\delta\delta'}(k, t) \,\mathrm{d}\mathbf{k}.
\end{align}
\end{widetext}

\begin{figure}[h]
	\includegraphics[width=1\linewidth]{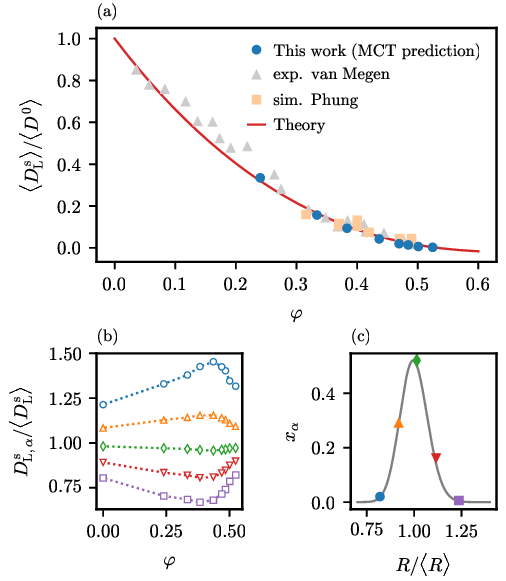}
	\caption{\label{fig:self-diffusion} (a) Average long-time self-diffusion coefficient $\langle D^\mathrm s_\mathrm L \rangle$ relative to the average Stokes-Einstein diffusion coefficient $\langle D^0 \rangle$ as predicted by the rescaled mode-coupling scheme compared to direct measurements by \protect\textcite{vanmegen1989} on a similar model system and Stokesian dynamics simulations by \protect\textcite{phung1993}. Additionally, a theoretical prediction based on parameterized simulation data in conjunction with the factorization approximation by \protect\textcite{medina-noyola1988} [Eq. 45 in Ref.~\protect\onlinecite{riest2015}] is given by the solid line. (b) Single-species long-time self-diffusion coefficient $D^\mathrm s_{\mathrm L, \alpha}$ relative to the average $\langle D^\mathrm s_\mathrm L \rangle$ in dependence on the volume fraction $\varphi$. The dashed lines are a guide to the eye. (c) Distribution of particle radii in relation to the mean, with each symbol in correspondence to (b). Additionally, the continuous Schulz-Flory distribution, which is approximated by the $n$-component mixture, is represented by the solid line.
	}
\end{figure}

The scheme for the self-dynamics relies on the short-time self-diffusion coefficients $D^s_\alpha$ which in the same spirit of an effective one-component treatment of the HIs are constructed to be $D^s_\alpha = H_\mathrm s D^0_\alpha$ where $H_\mathrm s$ is the self-part of the one-component hydrodynamic function [Eq.~\eqref{eqn:hydrodynamic-function}]. The rescaling procedure from \textcite{amokrane2017} is again incorporated by calculating the density-weighted partial direct correlation functions $C_{\alpha\beta}(Q)$ with the scaled volume fraction $\varphi^*$, exactly using the same values as for the collective dynamics. From the self-ISFs, we compute long-time self-diffusion coefficients $D^\mathrm s_{\mathrm L, \alpha}$ from the long-time, low-wave-vector limit of Eq.~\eqref{eqn:mct-self}, as described, e.g., in Ref.~\onlinecite{flenner2005}. Fig.~\ref{fig:self-diffusion}(a) shows the MCT prediction for the average long-time self-diffusion coefficient $\langle D^\mathrm s_\mathrm L \rangle_\alpha$ along with outcomes from independent computer simulations \cite{phung1993} and experiments on a similar model system \cite{vanmegen1986, vanmegen1989}. This is in accordance with a theoretical prediction based on the HI rescaling proposed in \textcite{medina-noyola1988}. Again, the values calculated via the mode-coupling scheme are in excellent agreement, both with literature data and theory which demonstrates the applicability of this rescaled MCT ansatz to consistently also describe tagged-particle motion, despite an additional layer of approximations within the theory. Furthermore, this especially emphasizes the intricate link between self-dynamics and cooperative effects.

On a final note, we want to point out a convenient by-product of the employed scheme: Since the procedure is based on a multi-component approach, automatically, species-resolved analyses become possible. In Fig.~\ref{fig:self-diffusion}(b), this is demonstrated by revealing what is essentially the distribution of long-time self-diffusion coefficients in the disperse suspension. Particularly intriguing is the evolution of the width of this distribution with increasing particle concentration. Compared to the width at $\varphi=0$ which is essentially just a representation of the particle size distribution [Fig.~\ref{fig:self-diffusion}(c)] expressed via the Stokes-Einstein relation, the distribution significantly broadens with increasing $\varphi$, up to a threshold value of $\varphi\approx0.45$, beyond which the relative diffusivities again converge towards the average value. The same convergence towards a shared time scale at high volume fractions was similarly observed recently in a quasi two-dimensional, binary hard sphere fluid \cite{thorneywork2017}. However, the exact mechanisms, especially the initial broadening of the distribution, are still not entirely clear.

\section{Conclusions}\label{scn:conclusions}
In this work, we describe quantitatively experimental collective density correlation functions $\phi_\mathrm M(Q, t)$ of liquid-like ordered model hard-sphere suspensions with a multi-component mode-coupling scheme based on the rescaled structure factor method proposed by \textcite{amokrane2017}. Hydrodynamic interactions are approximately incorporated using Medina-Noyola's \cite{medina-noyola1988} short-time factorization. Based on these results, long-time self-diffusion coefficients and zero-shear viscosities are derived from the collective dynamics without further adjustable parameters. Both quantities excellently agree with independent experiments and established theoretical predictions from the literature. Altogether, this demonstrates mode-coupling theory's capability as a tool for the consistent and quantitatively correct characterization of experimentally observed collective dynamics and transport properties in dense colloidal dispersions.

As an obvious downside, the first-principles character of MCT is lost within our approach due to the introduction of an empirical parameter without real physical significance. However, as already noted by \textcite{amokrane2019}, such a treatment is, beyond just being a practical solution for a complicated problem, certainly meaningful. Results such as these emphasize the immense potential of MCT if eventually, systematic advancements of the theory are indeed realized. A promising roadmap for this challenge was just recently laid out by \textcite{pihlajamaa2023}.

Additional investigations are required to further test the limitations of the here employed scheme. We restricted ourselves to the analysis of liquid-like or only mildly supercooled suspensions with volume fractions $\varphi\leq0.52$. A natural extension is thus the investigation of the more deeply supercooled and also the glass-like regime which probably requires the use of samples with broader size distributions to prevent an eventual crystallization even over long observation times \cite{ninarello2017}. If the here employed multi-component treatment is generally applicable to such very disperse systems is in itself a question of interest. Equally important is testing the predictions of such schemes for a more extensive set of dynamical observables, e.g., by probing the full, wave-vector-dependent self-dynamics or frequency-dependent viscoelastic properties.

Undoubtedly also desired is the application to model systems beyond hard spheres in bulk, with just a few examples being particles with long-range repulsive \cite{banchio2018} or competing interactions \cite{klix2010}, self-propelled particles \cite{reichert2021, debets2023} or systems in confinement \cite{jung2023}. From a technical perspective, this involves additional efforts to further develop methods for the fast evaluation of partial structure factors in multi-component systems, for example via advanced integral-equation schemes \cite{kalyuzhnyi2020}, and of partial hydrodynamic functions, where the only reasonably applicable analytic multi-component framework is based on Rotne-Prager-Yamakawa hydrodynamics, only suitable for dilute systems \cite{mcphie2007}.

\section*{Author Declarations}
\subsection*{Conflict of Interest}
The authors have no conflicts to disclose.

\end{document}